\documentclass{emulateapj}

\usepackage{xspace}
\usepackage{graphicx}
\usepackage{rotating}
\usepackage{natbib}
\usepackage{apjfonts}
\def\errtwo#1#2#3{$#1^{+#2}_{-#3}$}

\newcommand\msun{\mathrm{M_\odot}}

\newcommand\zth{$0^\mathrm{th}$}

\newcommand\chandra{\textsl{Chandra}\xspace}

\newcommand\integral{\textsl{INTEGRAL}\xspace}
\newcommand\isis{{\tt ISIS}\xspace}
\newcommand\iraf{{\tt IRAF}\xspace}

\newcommand\miriad{{\tt MIRIAD}\xspace}

\newcommand\mysou{IGR~J17177$-$3656}
\newcommand\osa{{\tt OSA}\xspace}

\newcommand\swift{\textsl{Swift}\xspace}

\newcommand\aproxgt{\mathrel{%
     \rlap{\raise 0.511ex \hbox{$>$}}{\lower 0.511ex \hbox{$\sim$}}}}
\newcommand\aproxlt{\mathrel{%
     \rlap{\raise 0.511ex \hbox{$<$}}{\lower 0.511ex \hbox{$\sim$}}}}

\slugcomment{Accepted 2011-06-25}

\shorttitle{IGR~J17177$-$3656}
\shortauthors{Paizis et al.}

\begin{document}

\title{Unveiling the nature of \mysou~with X-ray, NIR and Radio observations}

\author{A. Paizis\altaffilmark{1}, 
M.~A. Nowak\altaffilmark{2}, 
J. Wilms\altaffilmark{3}, 
S. Chaty\altaffilmark{4},
S. Corbel\altaffilmark{4},
J. Rodriguez\altaffilmark{4},
M. Del Santo\altaffilmark{5},
P. Ubertini\altaffilmark{5} 
and
R. Chini\altaffilmark{6,7}}

\altaffiltext{1}{Istituto Nazionale di Astrofisica, INAF-IASF, Via Bassini 15,
20133 Milano, Italy; ada@iasf-milano.inaf.it}
\altaffiltext{2}{Massachusetts Institute of
  Technology, Kavli Institute for Astrophysics, Cambridge, MA 02139,
  USA; mnowak@space.mit.edu}
\altaffiltext{3}{Dr.~Karl Remeis-Sternwarte and Erlangen Centre for
  Astroparticle Physics, Universit\"at Erlangen-N\"urnberg,
  Sternwartstr.~7, 96049 Bamberg, Germany}
\altaffiltext{4}{AIM  - Astrophysique, Instrumentation et Mod\'elisation 
(UMR-E 9005 CEA/DSM-CNRS-Universit\'e Paris Diderot)
Irfu/Service d'Astrophysique, Centre de Saclay 
FR-91191 Gif-sur-Yvette Cedex, France}
\altaffiltext{5}{IASF Roma - INAF, Via del Fosso del Cavaliere 100, 00133 Roma, Italy}
\altaffiltext{6}{Astronomisches Institut, Ruhr-Universit\"at Bochum, Universit\"atsstra\ss{}e 150, D-44780 Bochum, Germany}
\altaffiltext{7}{Instituto de Astronom\'ia, Universidad Cat\'olica del Norte, Avenida Angamos 0610, Casilla 1280,
Antofagasta, Chile}

\begin{abstract}
We report on the first broad-band (1--200\,keV) simultaneous \chandra-\integral observations of the recently discovered hard X-ray transient 
\mysou~that took place  on 2011, March 22, about two weeks after the source discovery. The source had an average absorbed 
1--200\,keV flux  of about 8$\mathrm{\times 10^\mathrm{-10}~erg~cm^\mathrm{-2}~s^\mathrm{-1}}$. 
We  extracted a precise X-ray position 
of \mysou, $\alpha_\mathrm{J2000}$=17$^\mathrm{h}$ 17$^\mathrm{m}$ 42$^\mathrm{s}$.62, 
\mbox{$\delta_\mathrm{J2000}$= $-$36$^{\circ}$ 56$^{\prime}$ 04.5$^{\prime \prime}$} 
(90\% uncertainty of 0.6$^{\prime\prime}$). We also report \swift, near infrared and quasi simultaneous radio follow-up 
observations.
With the multi-wavelength information at hand, we propose \mysou~is a low-mass X-ray binary, seen at high inclination, probably hosting a black hole.
\end{abstract}


\section{Introduction}\label{sec:intro}
\setcounter{footnote}{0}

On 2011 March 15 (MJD 55635), \integral discovered the new hard X-ray transient IGR~J17177$-$3656 \citep{frankowski11}. 
The IBIS/ISGRI  spectrum (20--200\,keV) could be well described by a power-law with photon index 1.8$\pm$0.3 
and a flux of about 20\,mCrab. The source was also marginally 
detected  in JEM-X  in the 10--20\,keV band (about 8\,mCrab) but was 
not detected in the 3--10\,keV band with a 3$\sigma$ upper limit of 5\,mCrab.

Following the \integral discovery, a \swift ToO was performed on 
2011 March 16 \citep{zhang11}. A refined \swift position was reported (90\% uncertainty of  2.1$^{\prime \prime}$),
consistent with the \integral one. 
A combined, though non simultaneous, \swift/XRT and \integral/IBIS/ISGRI spectral fit with an absorbed power-law 
model yielded $N_\mathrm{H}$=\errtwo{3.9}{0.5}{0.4}$ \times 10^\mathrm{22}~\mathrm{cm^\mathrm{-2}}$  and photon 
index $\Gamma$=1.5$\pm$0.2 indicating a hard state with additional absorbing column density intrinsic to the source \citep{zhang11}.
Based on the \swift position and uncertainty, a search through the VizieR database resulted only in one match: 
2MASS~J17174269$-$3656039 (K=12.9) located at about 1.3$^{\prime \prime}$ from the \swift position of \cite{zhang11}.\\

Our \chandra ToO was performed on 2011 March 22, and
thanks to the excellent \chandra location accuracy, an X-ray  source position  with a 0.6$^{\prime \prime}$ (90\%) uncertainty was reported \citep{paizis11}. The new \chandra-based position, not consistent with the \swift  one 
($\sim$4.8$^{\prime \prime}$ away, but see Section~\ref{sec:swiftresults}), probably  ruled out the 2MASS association (1.05$^{\prime \prime}$ away) and  allowed the scientific community as well as members of our team to search for NIR and radio counterparts \citep[this paper;][]{corbel11,torres11,rojas11}. \\
The nature of \mysou~is still to be unveiled and in this paper we describe our multi-wavelength campaign and 
contribution to its investigation.

\section{Multi-wavelength observations}
We present a detailed broad-band analysis of the recently discovered \mysou\footnote{See http://irfu.cea.fr/Sap/IGR-Sources/ for a regularly updated list of \integral sources} using our \chandra ToO data and simultaneous 
higher energy \integral ones. To obtain an overall view of the behavior of \mysou, we  also have analyzed  available \swift and  \integral
  monitoring data relevant to our investigation.
 Given our \chandra position, we triggered Near Infrared (NIR) observations to look for possible counterparts to \mysou. These data
 are included in this paper, as well as the quasi-simultaneous radio observation reported by \cite{corbel11}.

\subsection{\chandra~data}\label{sec:chandra}

We observed \mysou~for 20\,ks with \chandra on 2011 March 22, from 
06:07:15 UT until 12:00:48 UT (MJD 55642, Observation ID 12452) with the
High Energy Transmission Grating Spectrometer, HETGS \citep{canizares00} collecting high resolution spectral information with the  High Energy Grating, HEG 0.8--10\,keV, 
and Medium Energy Grating, MEG 0.4--8.0\,keV.
The data were analyzed in a standard manner, using the CIAO version 4.3 software 
package and \chandra CALDB version 4.4.2. The spectra were analyzed with 
the \isis analysis system, version 1.6.1 \citep{houck02}. 

\subsection{\integral~data}\label{sec:integral}
Starting from its discovery, \mysou~has been in the \integral/IBIS \citep{ubertini03} field of view during the Galactic
bulge monitoring\footnote{http://isdc.unige.ch/Science/BULGE/} \citep{kuulkers07},  as well as during the inner disk (\emph{l} < 0)
observations. A complete study of these \integral data is out of the scope of this paper. We will focus on the simultaneous \chandra/HETGS-\integral/IBIS
set, in the first simultaneous broad-band study of \mysou. To properly understand the  long-term hard X-ray behavior of the source (> 20\,keV) however, 
we have analyzed the IBIS/ISGRI data \citep[20--600\,keV;][]{lebrun03} starting from revolution 1028 (March 15, 2011, 09:23:08 UT, MJD 55635.39) to revolution 1032 (March 28, 2011, 22:16:46 UT, MJD 55648.93), spanning a
period of about 14 days. 
A standard analysis using version 9.0 of the Off-line Scientific Analysis (\osa) software
was performed.

\subsection{\swift~data}\label{sec:swift}
The 2011 outburst of IGR~J17177$-$3656 was monitored with a total of
eight daily pointed observations with the \swift X-Ray
Reflecting Telescope \citep{burrows05,gehrels04} between 2011 March 16
and 2011 March 23. Exposure times varied between 2\,ks and 8\,ks.
The data were processed using the newest version of the Swift data
analysis pipelines in HEASOFT 6.10. After re-processing the data with
\texttt{xrtpipeline}, spectra for were extracted with \texttt{xselect}
using XRT grades 0 through 12. We used the newest XRT response matrix
available in CALDB. Ancillary response matrices were generated using
\texttt{xrtmkarf}, taking vignetting effects into account.

In addition to these data, IGR J17177$-$3656 was in the field of view of
a 560\,s long XRT pointing taken on 2011 January 27, but it was not
detected during that observation.

\subsection{Radio data}\label{sec:radio}
We conducted  continuum radio observations with the Australia Telescope Compact Array (ATCA) located in Narrabri, New South Wales, Australia. The ATCA synthesis telescope is an east-west array consisting of six 22 m antennas. The ATCA uses orthogonal polarized feeds and records full Stokes parameters. We carried out the observations with the upgraded Compact Array Broadband Backend (CABB), which  provides a new broadband backend system for the ATCA and increases the maximum bandwidth from 128\,MHz to 2\,GHz.\\
The ATCA observations were conducted on the same day as the  \chandra observations, 2011 March 22, at two frequency bands simultaneously, with central frequencies at 5.5\,GHz and 9\,GHz. The ATCA was in the intermediate 1.5A configuration. A total observation time of 0.66\,hr on source was obtained for \mysou. The amplitude and band-pass calibrator was PKS~1934$-$638, and the antenna's gain and phase calibration, as well as the polarization leakage, were derived from regular observations of the nearby ($\sim$2.9$^{\circ}$ away) calibrator PMN~1714$-$336.  The editing, calibration, Fourier transformation, deconvolution, and image analysis were performed using the \miriad software package  \citep{sault95, sault10}. The cleaning process was carried out using a combination of multi-frequency \citep{sault94} and standard clean algorithms.

\subsection{Near-infrared  data}\label{sec:NIR}
Near-infrared $K_\mathrm{s}$ observations were performed at the
Universit\"atssternwarte Bochum near Cerro Armazones in the Chilean Atacama
desert. We used the 80 cm IRIS telescope equipped with a 1024 $\times$ 1024 pixels
HAWAII-1 detector array \citep{hodapp11}. The observational sequence
consisted of eight exposures between MJD 55647.38 and 55647.40 (2011 March 27, 09:10:50 < UT < 09:38:36) each comprising conventional dithering and
chopping patterns to allow subtraction of the bright NIR sky. The total
on-source integration time was 400\,s. Data reduction involved standard
\iraf procedures; astrometry and photometric calibration were achieved via
1958 sources from the 2MASS archive.

\section{Results}\label{sec:results}
\begin{figure}
\epsscale{1.2} \plotone{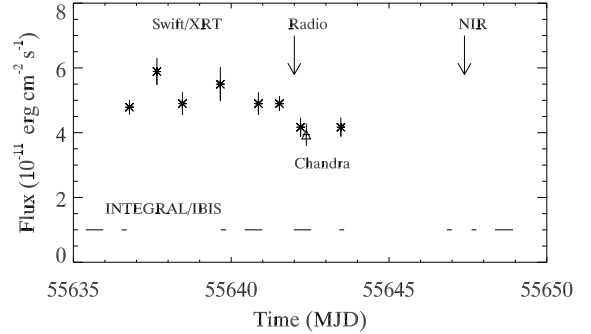}
\caption{\mysou~\swift and \chandra (absorbed) flux evolution in the 2--8\,keV range.  For clarity the \integral coverage of the source is also shown, marked as horizontal lines, together with our NIR and radio follow-up observations, 
as  discussed in this paper.}
 \label{fig:lcr_all}
\end{figure}

An overview of all the data treated in this paper  is shown in 
Fig.~\ref{fig:lcr_all}. The absorbed 2--8\,keV flux evolution of \mysou~as seen by \swift is shown together with  our \chandra average data point\footnote{The absorbed flux is shown, which we believe to be a more solid indication of the source state given its heavy intrinsic and variable absorbing column density.}. The \integral coverage 
of the source is marked as horizontal lines. Finally, the time of our NIR observation and of the radio observation included in this work are also shown.

\subsection{\chandra position, variability and spectra}\label{sec:chandraresults}
We extracted the X-ray position of IGR~J17177$-$3656 from the \zth-order image
obtaining  $\alpha_\mathrm{J2000}$=17$^\mathrm{h}$ 17$^\mathrm{m}$ 42$^\mathrm{s}$.62, 
\mbox{$\delta_\mathrm{J2000}$= $-$36$^{\circ}$ 56$^{\prime}$ 04.5$^{\prime \prime}$} \citep{paizis11}.
Given the brightness of the source, the statistical error is smaller than the 
absolute position accuracy of \chandra, 0.6$^{\prime \prime}$ at 90\% 
uncertainty\footnote{http://cxc.harvard.edu/cal/ASPECT/celmon/}. Therefore we attribute to the position 
found a 90\% uncertainty  of 0.6$^{\prime \prime}$.\\

In \cite{paizis11} we  performed a simple phenomenological fit to the MEG and HEG first order grating spectra. The \chandra total (about 20\,ks) spectrum was consistent with an absorbed power-law with a column density
of approximately $N_\mathrm{H}$=(5.9 $\pm$1.5$)\times 10^\mathrm{22}~\mathrm{cm^\mathrm{-2}}$ (90\% confidence) and 
photon index $\Gamma$=1.2$\pm$0.4 (90\% confidence) with the source at about a 3\,mCrab flux level in the 2--10\,keV range. The \chandra spectrum alone is very hard but this is likely due to the narrow energy range in which the 
fit is performed, which is strongly absorbed at lower energies by the heavy intrinsic absorbing column density present in the source.\\
A detailed analysis of the source light-curve, with 200\,s time bins, showed  variability, hence we 
cut the data in three, non-contiguous, intervals: a 2--8\,keV rate < 40 counts/bin 
(\chandra-Low, 4.6\,ks), between 40--60 counts/bin (\chandra-Mid, 4.5\,ks), and > 60 counts/bin (\chandra-High, 10.4\,ks).  
 
\begin{figure}
\epsscale{1} \plotone{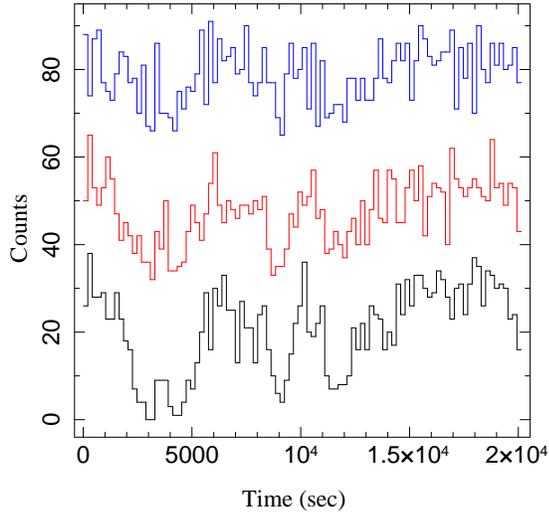}
\caption{200 s light-curves in 3 energy bands: 2--3.7\,keV (black), 3.7--4.7\,keV (red) and 
 4.7--8\,keV (blue). For visualization purposes, the middle  (red) and high band (blue) are offset by 30 and 60 counts  with respect to the  low (black) band.}
 \label{fig:lcurve}
\end{figure}

\begin{figure}
\epsscale{1} \plotone{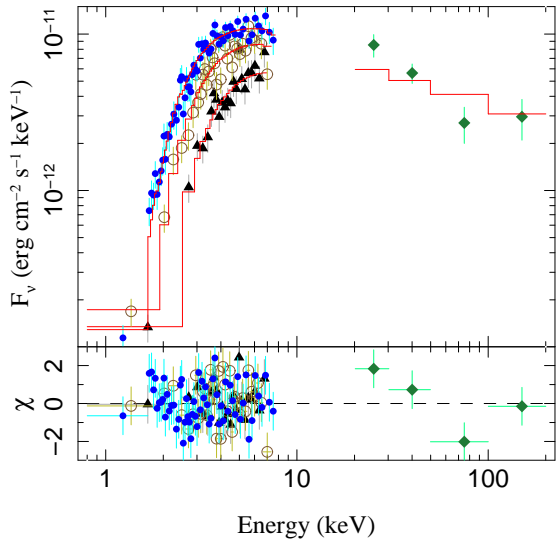}
\caption{Spectral model of Table~\ref{tab:efits}.}
 \label{fig:spectra}
\end{figure}

\begin{figure}
\epsscale{1} \plotone{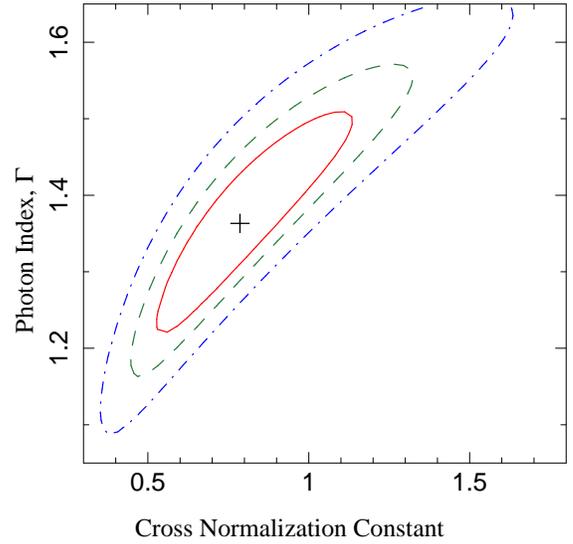}
\caption{Contours of IBIS/ISGRI normalization constant vs. photon
  index. (68\%, 90\%, and 99\% confidence for two parameters).}
 \label{fig:contour}
\end{figure}

\begin{deluxetable*}{lccccccr}
\setlength{\tabcolsep}{0.03in} 
\tabletypesize{\footnotesize}    
\tablewidth{0pt} 
\tablecaption{Fits to IGR~J17177$-$3656 Spectra: $C_\mathrm{cn}$*(TBnew(\textsl{ISM})*((TBnew(\textsl{local})+$C_\mathrm{f}$)*powerlaw))\label{tab:efits}}
\tablehead{
     \colhead{Instrument}
   & \colhead{$N_\mathrm{H}$}
   & \colhead{$C_\mathrm{f}$}
   & \colhead{$\Gamma$} 
   & \colhead{2--8 keV Flux\tablenotemark{a}}
   & \colhead{20--200 keV Flux\tablenotemark{b}}
   & \colhead{$C_\mathrm{cn}$}
   & \colhead{$\chi^2/$DoF}
          \\                               
   & ($10^\mathrm{22}~\mathrm{cm^\mathrm{-2}}$) & & 
   & ($\mathrm{\times 10^{-11}~erg~cm^{-2}~s^{-1}}$)
   & ($\mathrm{\times 10^{-10}~erg~cm^{-2}~s^{-1}}$)
         }
\startdata
   \chandra -- Low
 & \errtwo{16.8}{3.5}{3.1}       
 & \errtwo{0.015}{0.009}{0.019}  
 & \errtwo{1.36}{0.16}{0.15}     
 & \errtwo{2.36}{0.19}{0.19}     
 & \nodata                       
 & \textsl{1}                    
 & 130.9/111                     
\\
\noalign{\vspace*{0.4mm}}
   \chandra -- Mid
 & \errtwo{10.5}{1.7}{1.5}       
 & \errtwo{0.013}{0.008}{0.010} 
 & \nodata                      
 & \errtwo{4.03}{0.23}{0.22}    
 & \nodata                      
 & \textsl{1}                   
 & \nodata                      
\\
\noalign{\vspace*{0.4mm}}
   \chandra -- High
 & \errtwo{7.0}{0.9}{0.8}       
 & \errtwo{0.00}{0.010}{0.000}  
 & \nodata                      
 & \errtwo{5.44}{0.20}{0.19}    
 & \nodata                      
 & \textsl{1}                   
 & \nodata                      
\\
\noalign{\vspace*{0.2mm}}
   IBIS/ISGRI
 & \nodata                     
 & \nodata                     
 & \nodata 
 & \nodata 
 & \errtwo{6.48}{0.94}{1.15} 
 & \errtwo{0.79}{0.50}{0.26}
   \tablecomments{Errors are 90\% confidence level for one 
     parameter.  $N_\mathrm{H}$ is the neutral column density \emph{local to
       the system}. The ISM interstellar column density has been fixed to
     $1.5\times10^\mathrm{22}~\mathrm{cm^{-2}}$. $C_\mathrm{f}$ is a
     dimensionless constant to allow for partial covering (covering
     fraction $\equiv (1+C_\mathrm{f})^{-1}$). $\Gamma$ is the
     powerlaw photon index, which was constrained to be the same for
     all datasets.  
     For the IBIS/ISGRI data, all model parameters
     have been tied to those of the \chandra High dataset.  We have
     further allowed for a cross-normalization constant,
     $C_\mathrm{cn}$, for the IBIS/ISGRI detector. (This
     cross-normalization also subsumes the intrinsic flux variation of
     the \chandra spectra.)}  \tablenotetext{a}{Absorbed flux, relative to the \chandra
     response.}  \tablenotetext{b}{Relative to the IBIS/ISGRI response, and
     includes the cross normalization constant.}
\end{deluxetable*}

For each of the three intervals, we extracted again 
the first order dispersed spectra ($m=\pm1$ for HEG and MEG) and to 
increase the signal-to-noise ratio, we merged the two HEG ($m=\pm1$) and MEG ($m=\pm1$) spectra into one 
combined spectrum, for a total of three spectra (one per rate level)\footnote{The \chandra \zth-order, that qualitatively verifies the source flux variability, was not used in the spectral analysis as it suffers from pile-up.}. 
Final binning, starting at 0.8\,keV, was chosen to have 
a signal to noise ratio higher than 5 and a minimum of 16 MEG channels per bin. The  fitting results 
of these three spectra are given in Section~\ref{sec:ibischadnraresults} together with the simultaneous \integral/IBIS one. \\
Figure~\ref{fig:lcurve} shows the 200\,s light-curve in three energy bands 2--3.7\,keV (black), 3.7--4.7\,keV (red) and 
 4.7--8\,keV (blue). The bands were chosen to have roughly equal
integrated counts  and for visualization purposes
 the light-curves are respectively offset by 30 and 60 counts for the middle (red) 
 and high (blue) band, with respect to the low (black) band. As can be seen, the variability is more evident in the 
 softer band, see, e.g., the large dip-like structure right before 5000\,s
where the flux becomes undetectable in one 200\,s bin. Generally the lower the total counts, the harder the source (due to, for example, increasing absorbing column density and/or spectral slope change).

\subsection{\integral non-variability and spectrum}\label{sec:ibisresults}
For all the \integral coverage shown in Fig.~\ref{fig:lcr_all}, we extracted images
in the 17--50\,keV energy band (one image per pointing) to search for possible source flares. 
Among the
121 pointings analyzed, only in three was the source detected at a single pointing level 
(detection significance > 5$\sigma$). This result shows that in a single pointing \mysou~is (at best) at the detection limit in
IBIS/ISGRI, corresponding to about 25\,mCrab in a 1\,ks pointing\footnote{The higher detection significance quoted in the discovery ATel \citep{frankowski11} is obtained by mosaicking different pointings.}. Hence for the spectral analysis we decided to extract the source spectrum  from the mosaics (where the detection significance is higher,  using the \emph{mosaic\_spec}
tool available in the \osa package), rather than on a single pointing basis. A second run was thus made to extract images in the 20.18--30.23, 30.23--49.6, 49.6--100.12, 100.12--200.17\,keV bands (the energy boundaries are given by the response matrix itself).
This spectral extraction was made \emph{only} for the data that were \emph{simultaneous} to the \chandra  2011 March 22 ToO ($\sim$20\,ks).

 Given the three luminosity levels detected in \chandra data (Sect.~\ref{sec:chandra}), we extracted three spectra also for IBIS/ISGRI, matching as closely as possible the non consecutive \chandra-Low, -Mid, -High times. This resulted in about 4.3\,ks for \integral-Low (as opposed to 4.6\,ks for \chandra-Low), about 4.3\,ks for \integral-Mid (4.5\,ks for \chandra-Mid) and about 9.6\,ks for \integral-High (10.4\,ks for \chandra-High). The small differences in final integration times are due to the fact that whereas \chandra has a continuous observation, \integral is dithering hence the slew times between re-pointing are lost.\\ 
 The three spectra do not show any variability, IBIS/ISGRI is not capable of disentangling the three states due to the short integration time and low statistics. Alternatively,  we are seeing the extension of the trend of Fig.~\ref{fig:lcurve}, i.e., as we move to harder X-rays, the source is less variable.
Indeed the three IBIS/ISGRI spectra have comparable normalization and slope 
($\Gamma$$_\mathrm{Low}$= \errtwo{1.8}{0.5}{1.4}, 
$\Gamma$$_\mathrm{Mid}$= \errtwo{1.9}{0.8}{1.1},
$\Gamma$$_\mathrm{High}$= \errtwo{1.8}{0.5}{0.6}). Furthermore, we note that the IBIS/ISGRI spectrum reported in the discovery ATel  of \mysou~by \cite{frankowski11} (2011 March 15) could also be well described by a power-law with photon index 1.8$\pm$0.3, hence the hard X-rays did not experience a significant evolution from March 15 to March 22.

\subsection{\chandra-\integral simultaneous spectra}\label{sec:ibischadnraresults}
Since \mysou~does not show any appreciable variability in the IBIS/ISGRI range (> 20\,keV), to increase statistics 
we extract the IBIS/ISGRI average spectrum that is shown in Fig.~\ref{fig:spectra}, together with the three \chandra spectra (Low, Mid, High). The spectral model used to fit the combined spectrum is shown in Table~\ref{tab:efits}. A cross-normalization constant ($C_\mathrm{cn}$) was allowed between \chandra-grating and \integral/IBIS and was set to 1 for \chandra and free for \integral. 
Beside the interstellar medium (ISM) absorbing column density fixed to $1.5\times10^\mathrm{22}~\mathrm{cm^{-2}}$, a clearly varying local column density is needed, from $7.0\times10^\mathrm{22}~\mathrm{cm^{-2}}$ in \chandra-High to $16.8\times10^\mathrm{22}~\mathrm{cm^{-2}}$ in \chandra-Low (higher absorption, lower \chandra rate). \\
In the fit we have used an improved model for the absorption of X-rays in the ISM by 
\cite{wilms00} (so-called {\tt tbabs}). Such a model results in higher column densities with respect to, e.g., the {\tt wabs} model by \cite{morrison83}. Indeed in the earlier ISM absorption models, the 
abundances assumed for the ISM were the solar ones, while more abundance measurements  outside 
the solar system showed that the
total gas plus dust ISM abundances are actually lower than
the solar abundances. Hence with this correction, a higher column density is needed for a given spectrum \citep[][and references therein]{wilms00}. For this reason, the absorbing column density of \cite{paizis11}, which was determined from solar abundances, differs from the one given here.\\
During the fitting process it  became clear that  variability of $N_\mathrm{H}$
alone would not be enough to explain the \chandra variability.
Specifically, we noted that the spectral models that let free the neutral column density alone left 
positive residuals in the 0.8--3\,keV band, especially during the most absorbed phase.
Though this soft excess could 
be modeled in a number of ways, utilizing a partial covering fraction seems 
a natural hypothesis.  These factors, although small 
(\chandra-Low $C_\mathrm{f}$=0.015 and \chandra-Mid $C_\mathrm{f}$=0.013) 
improved the  $\chi^2$ by 12.
$C_\mathrm{f}$ was not applied to IBIS/ISGRI (> 20\,keV).

Given that the spectral slope does not show dramatic changes neither in the
already discussed \integral/IBIS spectra, nor in the \swift ones (see next
Section), in our fit the power-law photon index $\Gamma$ was constrained to be the same for all data-sets, letting the soft X-ray flux variability be explained by intrinsic $N_\mathrm{H}$ and (slight) covering.  Fig.~\ref{fig:contour} shows that care has to be taken for the final value of the spectral slope since it clearly depends on the \chandra-\integral cross-normalization. We note that trying to describe the broad-band data with a variable slope and fixed column density led to a very hard spectral index, $\Gamma$ down to 0.7, and cross-normalization $C_\mathrm{cn}$=0.5, i.e. the slope appeared to be too hard for IBIS/ISGRI, and its amplitude had to be halved to make the fit.
Although we cannot rule out  a contribution of the spectral slope to the overall variability, we assume the simplest model (fixed $\Gamma$, variable local $N_\mathrm{H}$). Finally we note that  a cut-off power-law is not strongly preferred ($\chi^2/$DoF=126.8/109), leading to an unconstrained cut-off energy of 
E$_\mathrm{c}$ > 34\,keV.

\subsection{\swift position and average spectrum}\label{sec:swiftresults}

The position of \mysou~determined from the summed image of the whole \swift
campaign is
 $\alpha_\mathrm{J2000}$=17$^\mathrm{h}$ 17$^\mathrm{m}$ 42$^\mathrm{s}$.4, $\delta_\mathrm{J2000}$=$-$36$^{\circ}$ 56$^{\prime}$ 03.6$^{\prime \prime}$  with a formal uncertainty
estimated to $3^{\prime \prime}$. We note that in the first \swift observation the
image of \mysou~was distorted by a chip gap, causing an offset in
the first \swift position reported by Zhang et al. (2011) and the
position reported here. Our inferred \chandra and \swift positions are consistent (2.8$^{\prime \prime}$ away).\\

\begin{figure}
\epsscale{1} \plotone{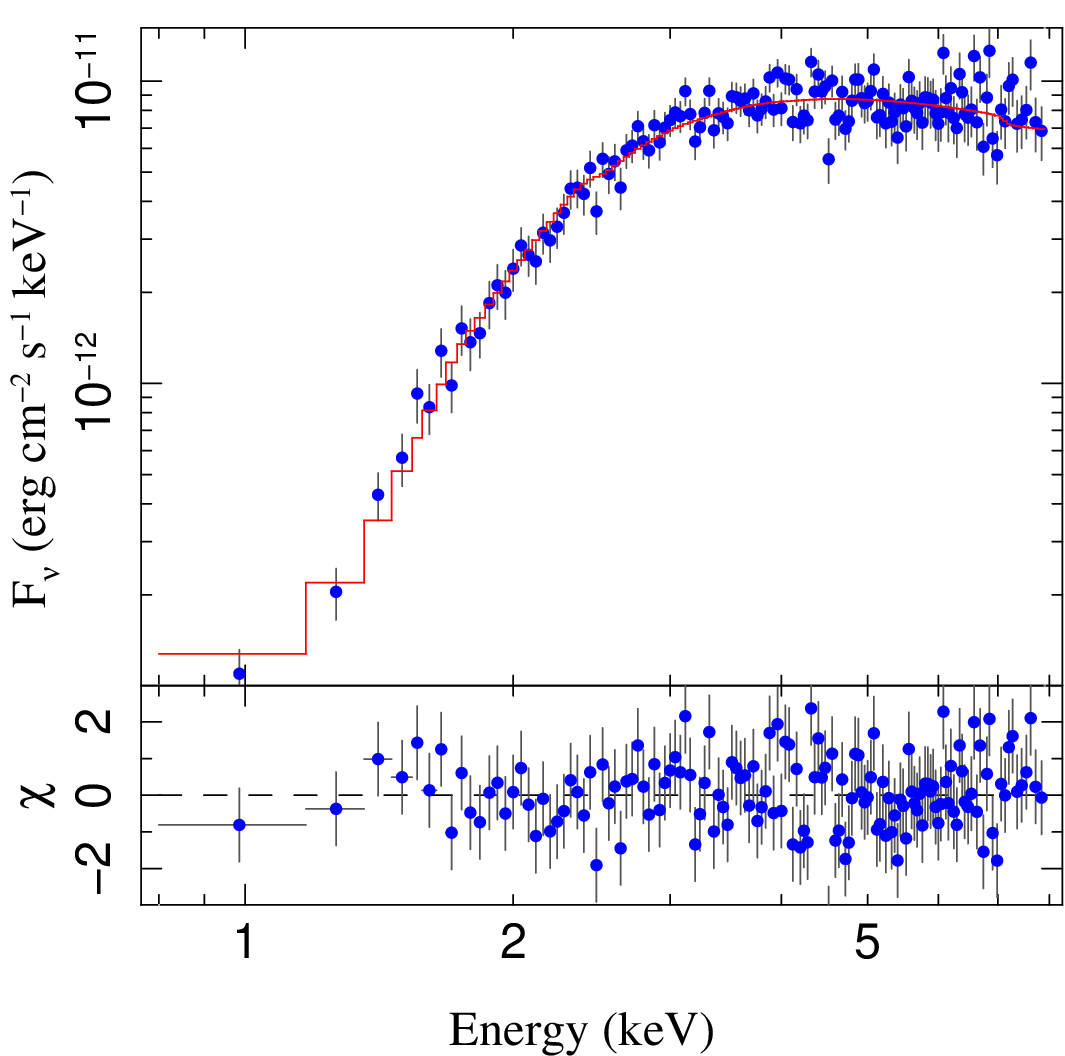}
\caption{\swift average spectrum of \mysou. See text for details.}
 \label{fig:swift}
\end{figure}

A spectral analysis of the observations shown in Fig~\ref{fig:lcr_all} has been performed. As for the \integral long-term data, a detailed analysis and discussion of the \swift results is out of the scope of the paper.
The spectra of the \swift/XRT data alone can be well described with an absorbed power-law throughout all the monitoring shown in Fig~\ref{fig:lcr_all}. The flux variability is less than a factor of two,  the spectral slope is constant within 90\% error bars, while  there seems to be an increase  in the absorbing column density in the last \swift point 
of Fig~\ref{fig:lcr_all}, just after our Chandra observation (i.e., the last point is not within 90\% error bars, but still within 99\%).
To have a global spectral view of the \swift results, since \mysou~does not show any significant variability in the \swift data, we co-added the single available spectra and fit the overall \swift spectrum (total exposure of about 16\,ks) using the same model structure of Section~\ref{sec:ibischadnraresults}.  We obtain a local absorbing column density 
$N_\mathrm{H}$=\errtwo{6.5}{0.6}{0.8}$\times10^\mathrm{22}~\mathrm{cm^{-2}}$, $\Gamma$=\errtwo{1.7}{0.2}{0.1}, covering factor $C_\mathrm{f}$=\errtwo{0.015}{0.007}{0.008} and absorbed 2--8\,keV flux F=(4.4$\pm$0.1)$\mathrm{\times10^\mathrm{-11}~erg~cm^\mathrm{-2}~s^\mathrm{-1}}$ (141.4/141 $\chi^2/$DoF). The average spectrum and fit is shown in Fig~\ref{fig:swift}. \\
The $N_\mathrm{H}$ value obtained from the average \swift spectrum is not 
compatible with  the average \chandra-\integral values and is 
in fact only compatible with the
least absorbed phases of the \chandra spectrum (\chandra-High).

in fact at its lowest end (\chandra-high). 
Similarly the \swift slope is 1.7 rather than around 1.4 for \chandra-\integral. We note however that the slope difference (and in turn $N_\mathrm{H}$) could be due to the fact that the \swift fit performed here is not coupled to the \integral data (as is \chandra). \cite{zhang11}, when fitting together (non simultaneous) \swift and \integral data, obtain a spectral slope $\Gamma$=1.5$\pm$0.2 (consistent with our result), albeit using a different absorption model than ours. \\
We emphasize that an accurate comparison cannot be done without a further investigation in the \swift, \chandra and \integral cross-normalization constants. Even within the \swift data alone, in order to establish a solid variability in the  absorbing column density, a deeper study is required, e.g. selecting different count-rate states from finer binning light-curves, extracting average spectra and comparing them,  as done for \chandra in this work.

\subsection{Radio}\label{sec:RadioRresults}

\begin{figure}
\includegraphics[width=0.9\linewidth,angle=270]{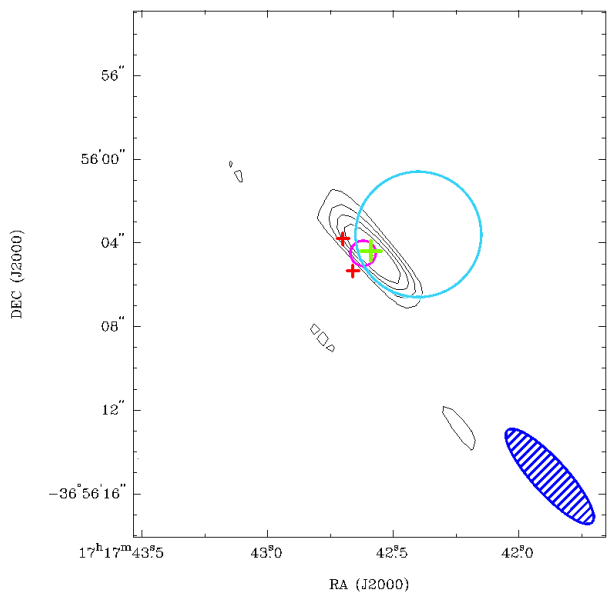}
\caption{Radio map using the combined ATCA observations of \mysou~at 5.5 and 9\,GHz (see text). The green cross
represents the fitted position of the ATCA counterpart, the purple circle is the 0.6$^{\prime \prime}$ \chandra
error, the light-blue one 
the 3$^{\prime \prime}$ \swift error, while the red crosses are the two NIR candidates discussed in the text \citep{torres11,rojas11}.}\label{fig:radio}
\end{figure}

Fig.~\ref{fig:radio} shows the obtained radio map using the combined ATCA observations of \mysou~at 5.5 and 9\,GHz.
Contours are at $-$3, 3, 4, 5,  and 6 times the r.m.s. noise level of 48\,$\mu$Jy/beam. 
The synthesized beam (in the lower right corner) is 6.0$^{\prime \prime}$ x 1.6$^{\prime \prime}$, with the major axis at a position angle of 42.3$^{\circ}$. 
The ATCA data indicate the presence of a single radio source within the \chandra X-ray error circle with
measured flux densities of 0.24$\pm$0.06\,mJy at 5.5\,GHz and 0.20$\pm$0.06\,mJy at 9\,GHz. The, poorly constrained, spectral index obtained is $\alpha$=$-0.37\pm$0.79. \\
The most accurate localization of the radio counterpart is obtained by combining the two frequency data-sets, giving a location of the radio source of $\alpha_\mathrm{J2000}$=17$^\mathrm{h}$ 17$^\mathrm{m}$ 42$^\mathrm{s}$.59, $\delta_\mathrm{J2000}$=
$-$36$^{\circ}$ 56$^{\prime}$ 04.4$^{\prime \prime}$ (0.5$^{\prime \prime}$, 1$\sigma$ positional uncertainty). Given its location
coincident with the \chandra one (0.37$^{\prime \prime}$ away, Fig.~\ref{fig:radio}), it likely corresponds to the radio counterpart of \mysou, though variability at radio frequencies would be needed to clearly establish this association.

\begin{figure}
\epsscale{1} \plotone{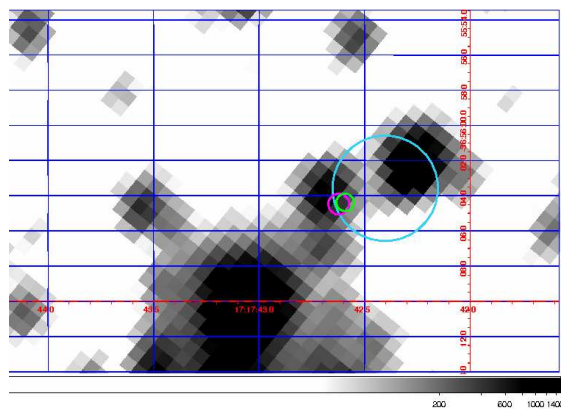}
\caption{Final averaged $K_\mathrm{s}$  image of the field of view around
\mysou. As in Fig.~\ref{fig:radio}, the green circle represents the ATCA position, the purple circle is the
\chandra error, while the light-blue one is obtained with \swift.}
 \label{fig:nir}
\end{figure}

\subsection{Near Infrared }\label{sec:NIRresults}

 The final averaged $K_\mathrm{s}$ image of the
field of view around \mysou~is shown in Fig.~\ref{fig:nir} and displays
sources down to $K_\mathrm{s} = 16.3$ mag.

The closest visible star to our
obtained \chandra position, visible in the figure, is  the already mentioned \mbox{2MASS~J17174269$-$3656039}  that seems to be blended 
with a weaker source. This blended doublet is present both in recent (2011 March 27, this paper and 2001 March 20, Torres et al., 2011) and archival NIR  maps (April and October 2010, Rojas et al., 2011). As already stated, the overall 90\% uncertainty circle of \chandra X-ray absolute position has a radius of 0.6$^{\prime \prime}$, while the 99\% limit on positional accuracy is 0.8$^{\prime \prime}$ with the worst case offset being 1.1. \mbox{2MASS~J17174269$-$3656039} is 1.05$^{\prime \prime}$ away from our position while the weaker source is 0.9$^{\prime \prime}$. These separations are very small and they may not rule out categorically the association to either source. However, assuming that the single radio source mentioned in Section~\ref{sec:RadioRresults}, compatible with \chandra position (0.37$^{\prime \prime}$ away), corresponds to the right counterpart, 
then the radio position would also tend to rule out  \mbox{2MASS~J17174269$-$3656039} and the faint infrared star. Furthermore, as already pointed out by \cite{torres11}
even taking the \swift absorption
 that is lower than \chandra-\integral one, would lead to ($\mathrm{J}-\mathrm{K}$)$\sim$6 that is higher than what is obtained for the blended sources \citep{torres11,rojas11}.   It is reasonable to conclude that no IR source is visible either in the 0.6$^{\prime \prime}$ Chandra position error indicated by the green circle, or in the 0.5$^{\prime \prime}$ ATCA circle (in red).

\section{Discussion}\label{sec:discussion}
In order to investigate the  nature of \mysou, we note that
\mysou~is located in the Galactic plane at (\emph{l},\emph{b})=(350$^{\circ}$.1, 0$^{\circ}$.51). 
Its X-ray emission together with its quasi-simultaneous radio emission, suggest that we are dealing with an X-ray binary, rather than with an AGN seen through the Galactic plane. 
From our least absorbed spectrum in Table~1, we obtain for the \emph{un-absorbed} fluxes, relative to \chandra normalization, F$_{\mathrm{3-9\,keV}}$=7.5$\times \mathrm{10^{-11}~erg~cm^{-2}~s^{-1}}$ and F$_{\mathrm{3-200\,keV}}$=1$\times \mathrm{10^{-9}~erg~cm^{-2}~s^{-1}}$.
Using the relation found by \cite{merloni03} that defines a "fundamental plane" 
in the three-dimensional (L$_{\mathrm{< 10\,keV}}$, L$_{\mathrm{R}}$, M) space, and our quasi-simultaneous X-ray (un-absorbed 3--9\,keV) and  radio emission, we see that we would need to place the source at about 100\,Mpc to obtain a 3--9\,keV luminosity of $\sim$$\mathrm{10^{44}~erg~s^{-1}}$, obtaining an estimated mass of about 4.4$\times$$10^\mathrm{4}$ $\msun$ which is too low for an AGN. Placed at 1\,Gpc, the source would have a luminosity as high as L$_\mathrm{X}$$\sim$$\mathrm{10^{46}~erg~s^{-1}}$, still not reaching $10^\mathrm{6}$ $\msun$ (4.7$\times$$10^\mathrm{5}$ $\msun$), which is extremely unlikely for an AGN scenario. \\
On the other hand, considering a binary at a distance of 8\,kpc would result in an un-absorbed luminosity  L$_{\mathrm{3-9\,keV}}$=5$\mathrm{\times 10^{35}~erg~s^{-1}}$ and L$_{\mathrm{3-200\,keV}}$=7$\mathrm{\times 10^{36}~erg~s^{-1}}$ which is compatible with an X-ray binary luminosity. \\
Although the \chandra observation occurred already in the declining phase of the outburst, its flux is comparable to the peak seen by \swift (see Fig.\ref{fig:lcr_all}), hence \mysou~likely belongs to the so-called class of Very Faint X-ray Transients (VFXT), i.e., transients showing outbursts with low peak luminosity 10$^\mathrm{34}$--10$^\mathrm{36}$ erg~s$^\mathrm{-1}$ in 2--10\,keV \citep{wijnands06}.\footnote{Indeed even considering the
\swift peak flux according to \cite{zhang11} (occurring on MJD 55636.8), we obtain a non absorbed 
F$_{\mathrm{2-10\,keV}}$=8.2$\mathrm{\times 10^{-11}~erg~cm^{-2}~s^{-1}}$, leading to  
L$_{\mathrm{2-10\,keV}}$=6$\mathrm{\times 10^{35}~erg~s^{-1}}$ at 8\,kpc.}\\
VFXT are believed to be the faintest known accretors, and are very likely a non-homogeneous class of 
sources. It is likely that most of them are
neutron stars and black holes  accreting matter from a low mass
companion \citep{wijnands06} and it has been found that a significant fraction of them
($\sim$1/3) have exhibited type-I X-ray bursts \citep[][and references therein]{delsanto2010}.
This is currently not the case for \mysou, and the black hole possibility, though more difficult to infer, is still open (see Section \ref{sec:radio}).

\subsection{LMXB or HMXB?}\label{sec:HorL}

 In order to speculate over the nature of the companion (hence high-mass X-ray binary, HMXB, versus Low-mass X-ray
 binary, LMXB), we have to consider the $K_\mathrm{s}$ limits obtained in quiescence, so that we are not contaminated
 by the disc X-ray to NIR re-processing, as it would be the case for an LMXB. The archival NIR maps studied by
 \cite{rojas11} show that no source was detected within either the \chandra  or the ATCA radio error circles in
 April-October 2010 maps, down to a limit of K$_\mathrm{s}$$\sim$18.0. Using the relation of \cite{prehdel95}, a visual extinction of A$_\mathrm{V}$=39.1 magnitudes is derived from $N_\mathrm{H}$=7$\mathrm{\times 10^{22}~cm^{-2}}$ (Table~\ref{tab:efits}). Assuming a spectral type supergiant Ib (B0.5) with M$_\mathrm{V}$ = $-$6.8 and $\mathrm{V}$-$\mathrm{K} = $-$0.75$ \citep{ducati01} (i.e., an HMXB), a distance of 8\,kpc, and using the extinction law by \cite{cardelli89}, A$_\mathrm{K}$/A$_\mathrm{V}$=0.114, we obtain an estimated observed K=12.92 that would have been detected in the maps by \cite{rojas11} (the NIR emission from an HMXB does not vary dramatically due to its bright stellar emission). Hence in the case of an HMXB hosting a supergiant, to reach the K$\sim$18.0 limit we would need to place the source at about 83\,kpc or more. Given our un-absorbed X-ray flux 
 (F$_{\mathrm{3-200\,keV}}$=1$\times \mathrm{10^{-9}~erg~cm^{-2}~s^{-1}}$), this would result in an X-ray luminosity of about L$_\mathrm{(3-200\,keV)}$$\sim$8$\times$10$^\mathrm{38}$ erg~s$^\mathrm{-1}$ from an HMXB placed outside the Galaxy. This scenario, though not impossible, is highly unlikely since such luminosities are too high for HMXBs with the compact object accreting stellar wind.  \\
 On the other hand, Be stars (representing the majority of companions in HMXB systems) are main sequence stars of spectral type between 09 and B2, with a spread in
absolute magnitude of nearly 1 mag \citep[see, e.g.,][]{allen84}. Be X-ray binaries, although dimmer than
supergiants, are usually bright NIR sources. The Be phenomenon itself, with the presence of a decretion
disk due to fast stellar rotation close to disruption, increases the luminosity of the source, at a
level of about K=0.25 mag \citep{dougherty94}. To give an example, a normal B0V star (absolute
magnitude M$_\mathrm{V}$=$-$4.1) with apparent K$\sim$18 mag and A$_\mathrm{V}$ = 39.1 mag would be at nearly 25\,kpc, and the disc
emission would have the effect of increasing its distance of nearly 3\,kpc, placing this star at nearly
28\,kpc. 
This is a substantial distance for instance in comparison with the faintest ($K_\mathrm{s}$=15.75) and furthest
(> 8.5\,kpc) known candidate BeXRB in the Galaxy, described in \cite{zurita08}. In general however, distances of HMXBs, both hosting Be and supergiant stars, are usually of the order of a few
kpc, due to their lower intrinsic X-ray luminosities than LMXBs, and they are distributed mainly
towards tangential directions of the Galactic arms \citep{grimm02}.\\
Furthermore, direct accretion through stellar wind --- common in HMXBs --- should prevent the formation of
strong and coherent jets, while the detection of radio emission from \mysou~suggests the
presence of ejection phenomena, and therefore an LMXB nature filling its Roche Lobe and forming an
accretion disk.\footnote{One could suppose the case of Cyg X$-$1, which exhibits both an accretion disk and stellar wind, but this
source is the only example known in our Galaxy of BH with Roche-lobe overflow in a supergiant X-ray
binary. In general, HMXBs strictly filling their Roche-Lobe are likely not observable, since the mass
transfer is highly unstable and the accretion should only last for a few thousand years. Instead, there
are HMXBs exhibiting Beginning Atmospheric Roche Lobe Overflow, where the massive star does not fill
its Roche Lobe, but the stellar wind follows the Lagrange equipotentials, and accumulates, forming an
accretion disk \citep{bhattacharya1991}. This
situation is more stable, but still rare due to the required configuration of stellar radius, orbital
distance and mass ratio, since we know only 3 such systems in total, hosting neutron stars: LMC~X$-$4,
Cen~X$-$3 and SMC~X$-$1.}\\

All these arguments suggest that the HMXB nature can be reasonably ruled out, pointing towards an LMXB
nature for \mysou. Indeed, assuming a spectral type main sequence K5 star with M$_\mathrm{V}$=7.3 and $\mathrm{V}$-$\mathrm{K}$=2.66 mag \citep{ducati01} (i.e., an LMXB) and a distance of 8\,kpc, we would obtain K$\sim$23 mag, compatible with the non-detection in \cite{rojas11}, and an X-ray luminosity of about L$_\mathrm{(3-200\,keV)}$=7$\times$10$^\mathrm{36}$ erg~s$^\mathrm{-1}$ that, though dim, is not  exceptional for the LMXB outburst luminosities.
 
 \subsection{A low-mass X-ray binary: geometry and compact object}\label{sec:radio}
 \begin{figure}
\includegraphics[width=0.9\linewidth]{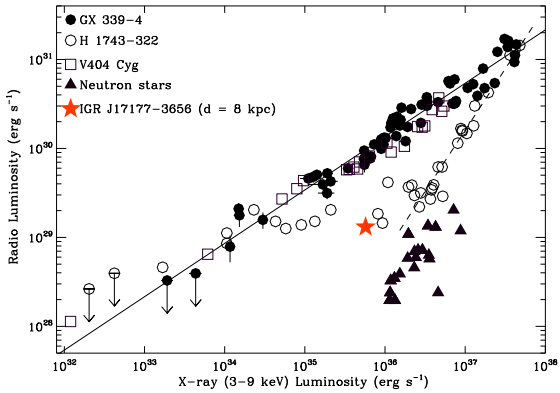}
\caption{Radio to X-ray luminosity for a sample of black hole and neutron star LMXBs. Adapted from \cite{coriat11}, see text for details.}\label{fig:radiotoX}
\end{figure}

Adding the local and Galactic (ISM) absorbing column densities of Table~1, we observe that we obtain 
a column density up to about ten times more  than the 
Galactic average value expected in the source direction, $\sim$1.3$\times$$10^\mathrm{22}$\,cm$^\mathrm{-2}$ \citep{kalberla05}.
 An LMXB does not have an important wind from the companion, that is generally an old  K-M type star, but if it is seen at a high inclination (> 60$^{\circ}$) it will appear as
 heavily absorbed due to the material  that gets into the line of sight (disc, blob of impact of the accreting matter onto the disc, the companion
 itself, etc). This is the case for the dipping, eclipsing or so-called ''coronal'' sources, i.e., sources that exhibit dips, total or partial eclipses in the soft X-ray spectra \citep[e.g. X1822$-$371, EXO~0748$-$676, 4U~1624$-$490;][]{parmar86,white82,xiang2009,diaz06}. \\
In our case, it is difficult to state if we are in the ''pure-dipper'' scenario
(inclination between 60 and 75 degrees) or at higher inclinations (eclipsing or coronal source).  We do not
see any total eclipse in our data (zero flux) but our soft X-ray observations are either continuous 20\,ks
(5.5\,hrs, \chandra) or quick snapshots that add up to about 15\,ks (\swift) which is a small part of a
typical LMXB orbital period (typically below 15--20\,hrs). For such a dim source, we would  need to integrate many orbital periods to
observe a  modulation  (dip, eclipse etc), but what is clear is that the environment we are observing is not
''clean'' (high $N_\mathrm{H}$) and it is irregular and patchy (the three rate levels in \chandra are not
contiguous in time). \\
We note that the non-variability in the IBIS/ISGRI spectrum together with the
absence of total eclipses (albeit the short coverage) could suggest that the Comptonizing region (responsible
for the photons above 20\,keV) would be extended and wider than the blocking material. Such ''coronal'' sources
are usually known as Accretion Disc Corona sources (ADC), high inclination systems   where the accretion disc
completely blocks the line of sight to the central source at all orbital phases, but that can still  be
observed because the corona allows X-rays to be scattered into the line of sight. If this is the case for \mysou, then 
the ''true'' X-ray flux could be higher. This is very similar to what obtained by \cite{xiang2009} on the high inclination LMXB,  so-called ''Big Dipper'', 4U~1624$-$490, based on a \chandra grating  observation over
the $\sim$76\,ks binary orbit of the source. The continuum spectrum could be modeled using a single $\Gamma$$\sim$2 power-law partially covered by a local absorber of column density $N_\mathrm{H}$$\sim$8$\mathrm{\times 10^{22}~cm^{-2}}$
besides the Galactic one, similarly to \mysou. 
Unlike \cite{xiang2009}, our data do not allow us to constrain discrete lines in the spectrum (due to the short exposure and dim source), but we can 
speculate that the scenario for \mysou~could be similar to the Big Dipper case for which \cite{xiang2009} have shown that  X-ray variations are predominantly driven by changes in obscuration, rather than intrinsic variation of the components.\\

LMXBs are  known to produce radio emission, be they hosting a black hole or a neutron star. Though black hole binaries are known to be more radio-loud, there are clearly some broad similarities in their X-radio coupling, such as the association between X-ray spectral states and the presence of radio emission \citep{fender06,corbel03,coriat11,migliari06,paizis06}. Radio emission can result
 either from a powerful compact jet in the hard state, or from relativistic discrete ejections at  spectral state transitions \citep[from hard intermediate to soft-intermedite state, e.g.,][]{fender06}.  
While  discrete radio ejections show a radio spectrum
with spectral index $\alpha<0$,  compact jets 
have a flat or slightly inverted spectrum (i.e. $\alpha\gtrsim0$). In our case, the poorly constrained  $\alpha$=$-0.37\pm$0.79 does not allow us to firmly establish the nature of the radio emission, however  since \mysou~ has not shown any important spectral transition but stayed in the hard state,  
it is likely that the radio emission is  associated with compact jets.\\
The quasi-simultaneous observation and detection of \mysou~with \chandra and ATCA is very important because besides leading to a constrained source position, it basically rules out the AGN and HMXB nature of the source. Indeed \mysou~fits
 well in the LMXB scenario also from the radio emission point of view (regardless of the inclination). 
 Figure~\ref{fig:radiotoX}, adapted from \cite{coriat11}, shows the radio luminosity against the 3--9\,keV un-absorbed luminosity for a sample of black hole candidate LMXBs in the hard state (H~1743$-$322, GX~339$-$4 and V404~Cyg) and atoll neutron star LMXBs in the island 
 state (Aql~X$-$1 and 4U~1728$-$34). The line is the fit to GX~339$-$4 from Corbel et al. (in preparation). 
 \mysou~is also shown in the plot. The assumed distance is 8\,kpc with an un-absorbed X-ray flux of 
 F$_\mathrm{(3-9\,keV)}$=7.5$\mathrm{\times 10^{-11}~erg~cm^{-2}~s^{-1}}$ (obtained from the least absorbed case) and a radio flux of F$_\mathrm{(radio)}$=0.2\,mJy.  \mysou~seems to follow the behavior of the black hole candidate LMXB H~1743$-$322 at low fluxes,
 when it starts to join the standard correlation of GX~339$-$4 \citep{corbel11,coriat11}. This could possibly suggest
 that \mysou, that appears also to be too radio-bright to be a neutron star LMXB, is a black hole candidate in the hard 
 state. We note however that we cannot firmly exclude the neutron star nature of the compact object based on
  the location of \mysou~in Fig.~\ref{fig:radiotoX} alone, since in the aforementioned ADC scenario the observed 
  flux (hence luminosity) could be underestimated. A higher ''true'' X-ray flux would indeed move the source 
  towards the neutron star populated area.

\acknowledgments 

We thank the \chandra team for their rapid response in
scheduling and delivering the observation, 
as well as the \integral Science Data Center for their quick and efficient 
sharing of \integral results. \\
Partly based on observations with \integral, an ESA project
with instruments and science data center funded by ESA member states, Czech Republic and Poland, and with the
participation of Russia and the USA.
This publication makes use of data products from the Two Micron All 
Sky Survey, which is a joint project of the University of Massachusetts and the 
Infrared Processing and Analysis Center/California Institute of Technology, 
funded by the National Aeronautics and Space Administration and the National 
Science Foundation. \iraf is distributed by the National Optical Astronomy Observatory, which is 
operated by the Association of Universities for Research in Astronomy (AURA) 
under cooperative agreement with the National Science Foundation. The Australia Telescope is funded by the Commonwealth of Australia for operation as a national Facility managed by CSIRO.\\
AP, MDS and PU acknowledge financial contribution from the agreement ASI-INAF I/009/10/0.
MDS acknowledges the grant from PRIN-INAF 2009 (PI: L. Sidoli).
This work was supported by NASA Grant GO1-12054X and partly supported by the Centre National d'Etudes 
Spatiales (CNES), based on observations obtained with MINE: the Multi-wavelength INTEGRAL NEtwork.
JW and SC acknowledge support from  European Community's Seventh Framework Programme (FP7/2007-2013) under grant
agreement number ITN 215212 "Black Hole Universe".
JW acknowledges support also from
the Bundesministerium f\"ur Wirtschaft und Technologie through Deutsches
Zentrum f\"ur Luft- und Raumfahrt Grant 50 OR 0801. \\
AP acknowledges Nicola La Palombara for the generous sharing of his laptop during an emergency, allowing 
us to trigger our \chandra observation in due time, and Lara Sidoli for useful discussion.

\newpage

\bibliographystyle{apj}
\bibliography{biblio}

\end{document}